\documentclass[%
 reprint,
 amsmath,amssymb,
 aps,
]{revtex4-2}

\usepackage{graphicx}
\usepackage{dcolumn}
\usepackage{bm}
\usepackage{ulem}[normalem]
\usepackage{xcolor}

\begin{document}

\title{
Inner Structure of Dark Matter Halos at High $z$ in Cosmological Models with Non-power-law Primordial Spectra\\
} 

\author{M.V. Tkachev}\email{mtkachev@asc.rssi.ru}
\author{S.V. Pilipenko}%
 \email{spilipenko@asc.rssi.ru}
\author{E.V. Mikheeva}
\email{helen@asc.rssi.ru}
\author{V.N. Lukash}
\email{lukash@asc.rssi.ru}
\affiliation{%
Astro Space Center of P.N. Lebedev Physical Institute, Moscow, Russia}



\date{\today}

\begin{abstract}
We consider three cosmological models with non-power-law spectra of primordial density perturbations and test them against $\Lambda$CDM in density profiles of dark matter halos. We found that, despite the significant difference in initial conditions, the mean density profiles of all models are still close to the Navarro-Frenk-White one, albeit with some dispersion. We demonstrate that the density profile slopes in the innermost part of halo have a significant evolution with $z$, which can be used to identify the cosmological model. We also present a toy model resulting in the appearance of core in the central part of dark matter halo.    
\end{abstract}

\maketitle


\section{\label{introduction}Introduction}
Recently non-trivial spectra of density perturbations have attracted considerable interest \cite{hirano15, Hutsi23, parashari23, padma23, tkachev23, 2023JCAP...04..011I, hirano24, 2024arXiv240214079R}. Such spectra are a significant extension of power-law spectrum of density perturbations, predicted in one-field inflationary models. Unlike to smooth power-law, they are capable to provide an additional enhancement at small scales, which can be related with Population III stars \cite{hirano15} or excess of high-$z$ galaxies \cite{tkachev23}. If an enhancement of a power spectrum is high,  primordial black holes can be born (see \cite{2023JCAP...04..011I}). 

In this paper we consider two kinds of primordial spectrum of density perturbations. The first one (model $gauss\_k15$) has the Gaussian bump and was studied in \cite{tkachev23} to clarify its impact on the halo mass function (HMF). The simplest way to produce a power-law spectrum with bump is to consider a single-field inflation with a kink in the potential (see \cite{star, inn} and more detailed consideration in the recent review \cite{inomata}). The specific shape of the spectrum can vary and depends on the inflationary model. Therefore, instead of analyzing inflationary potentials, one can use a phenomenological approach, assuming a single general feature added to the power-law spectrum. As in our previous work \cite{tkachev23} we choose as such a feature a Gaussian.

Another method to enhance the power spectrum  of density perturbations is to assume a double, or multiple as a more common case, field inflation. However, in the latter case it is still assumed to observe a rather short interval of inflationary stage compared to the total duration, so it may be unobservable, and hence such multistage inflation will be perceived as double one. In this kind of inflation the primordial spectrum of density perturbations has a minimum at some scale (see, for example, fig.~1 in \cite{lm1}). Such spectrum has a ''red'' wing on large scales and ''blue'' one at small scales, and precise forms of wings depend on the total inflaton potential. Following to notification proposed in \cite{hirano15} we refer these models as ''blue tilted'' ones.

Earlier, we have been studying the effect of power spectrum with a bump on the HMF, hereafter we concentrate on its influence on the density profile of dark matter (DM) halos.
As the \textit{gauss\_k15} model is in good agreement with observational data on high-$z$ galaxies found by \textit{James Webb Space Telescope (JWST)}, we use it as the first choice and add two blue tilted models with different tilts and scales of a minimum. 
Therefore, in paper we test cosmological models with different deviations at sub-Mpc scale in density profile and study their influence on the structure of the innermost part of dark matter halos. Profiles of dark matter halos for the non-standard spectra have been studied several times: \cite{Knollmann_2008} analyzed scale-free spectrum with varying power law slope. A number of studies consider warm dark matter with the power spectrum cutoff (see, e.g. \cite{bose17}). In all these cases the density profiles retain cusps. To our knowledge, there were no studies of the inner structure of halos for the two variants we consider here: the bumpy spectrum and the blue tilted one. Previously we have shown that the complex interaction of the small and large scale perturbations can result in the significant change of halo profiles \cite{pilipenko12}. This motivates us to check if the non-standard shape of the spectrum can also lead to similar effects or some other deviations of the density profiles of halos from the expected in the $\Lambda$CDM cosmology.

The shape of density profile of dark matter halos has been widely discussed over 30 years, and it is known as \textit{the cusp problem}. It can be summarized as a discrepancy between observations and simulations in the inner slope of radial density profile: simulations predict a profile with the behavior close to $r^{-1}$ at small radii (cusp) while observations show the existence of constant density cores at least in some DM-dominated galaxies. More details can be found in, e.g., \cite{deBlok}. A number of solutions to the cusp problem have been proposed, which can be divided into four classes: errors in the interpretation of observations, errors in simulations, baryonic physics, and the manifestation of the ``new'' physics. All the proposed solutions have some drawbacks, so there is no final consensus on the solution of the cusp problem yet \cite{CC_solutions}.

Numerical N-body simulations in cosmology are widely used and have been tested for convergence many times \cite{power03,trenti10,klypin13,Baushev}, however there are still some nuances. First, all the simulations cut the initial power spectrum of density perturbations at some scale (corresponding to the Nyquist wavenumber). Since in the standard WIMP $\Lambda$CDM model the spectrum does not drop in amplitude up to very small scales (the free-streaming scale for 100~GeV WIMP is about $10^3$ AU \cite{2008PhRvD..77b3528B}), the dark matter actually should be very clumpy at small scales.
The lack of resolution results in numerical errors close to the Nyquist wavenumber \cite{joyce09}.
Also, the nonlinear evolution of small-scale perturbations missed by numerical simulations should result in some kind of heating of dark matter particles, or decreasing its mean phase space density. This ''heating'' can be understood in the context of Lynden-Bell's definition of entropy \cite{lyndenbell}, where it corresponds to a decrease in the "fine-grained" phase space density of dark matter particles. Since a cusp is a region of high phase space density (low entropy), the missing small-scale power may facilitate the formation of cuspy profiles. This has been proposed in \cite{DLM2007, LM2010, DLM2008}.

Several attempts to simulate halo formation from the free-streaming scale have been made \cite{ishiyama14,delos23}, but this required peculiar initial conditions: in \cite{ishiyama14} the box size was limited to 400~pc while in \cite{delos23} the simulated halo was selected in a void many times below the mean density. 
So these simulations do not fully answer the question of how the perturbations at very small scales (free streaming) interact with perturbations at much larger scales (galaxy scale), because in these simulations larger scale waves were significantly damped by the technically limited choice of the initial conditions.
On the other hand, high resolution simulations show that with the increase of resolution (the number of particles per halo) the profile changes from the Navarro-Frenk-White (NFW) one \cite{NFW} to the Einasto profile \cite{Einasto} with somewhat shallower density in the center. This is expected from the theory proposed in \cite{DLM2007}. However, the Einasto profile still cannot explain the observations of cored galaxies, since the size of the core in them is much larger than the region in Einasto profile with the shallow density slope.

The second source of possible errors in simulations is the fact that the cosmological codes give an approximate solution for the N-body problem (see, e.g., \cite{zhan06}). It has been noted that this may result in the cusp being an attractor solution of an approximate N-body \cite{baushev15,baushev17,Baushev}. Also simulations are prone to artificial disruption of satellites which may bring additional DM particles to halo centers in simulations \cite{Bosch}.

While the propositions of cusp formation by \cite{Baushev} and \cite{Bosch} are hard to check or improve, the entropy method, proposed in \cite{DLM2007}, predicts that by introducing additional power on small scales one can compensate the deficit of small scale perturbations in simulations. In this Paper we aim at testing this prediction by adding this power. This can be done in several ways, either by changing the power spectrum by increasing the small scale power, or by introducing small scale random velocities.

So we have double interest in simulating universes with bumpy or tilted power spectra. First, such spectra may be physical as they can arise in various inflation models. If such models produce significant amount of cored halos, the density profile can be used as a test for these inflationary models. Second, the addition of small scale power allows us to check the ideas proposed by \cite{DLM2007, LM2010, DLM2008} and check if the missing small scale power is promoting cusp formation.
\section{\label{description}Models description}
In order to investigate the impact of the spectrum modification on the evolution and the inner structure of dark matter halos, we employed power spectra constructed as the product of the standard $\Lambda$CDM spectrum and a certain transfer function. In case of the Gaussian bump it remains the same as in our previous work \cite{tkachev23}:
\begin{equation}
    T(k) = 1 + A \cdot \exp \left( -\frac{(\log(k)-\log(k_0))^2}{\sigma_k^2} \right), 
    \label{bumps}
\end{equation}
where $k$ is a wave number, $A$, $k_0$, and $\sigma_k$ are bump parameters and we assume a value of $\sigma_k=0.1$. As before, we would like to point out that the shape (\ref{bumps}) is not predicted directly by simple modifications of the inflation model (see, e.g., \cite{2023JCAP...04..011I}). We consider the Gaussian shape as a simple approximation of the peak shape in various models.

\begin{figure}
\includegraphics[width=1.0\linewidth]{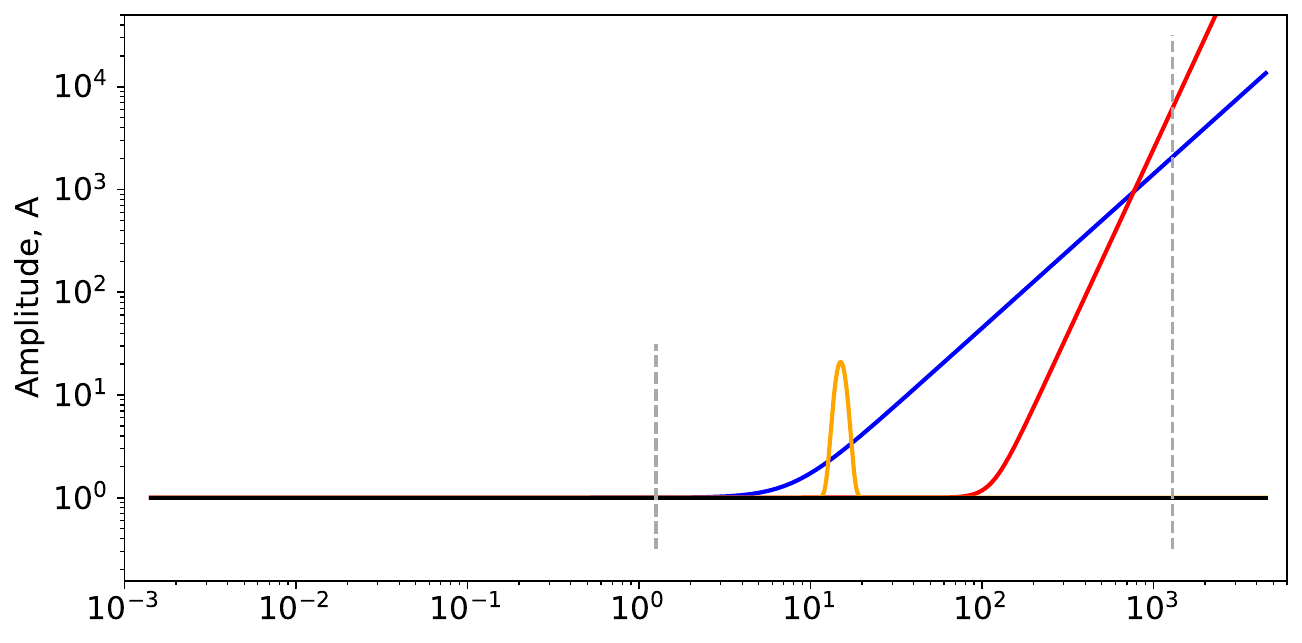}
\includegraphics[width=1.0\linewidth]{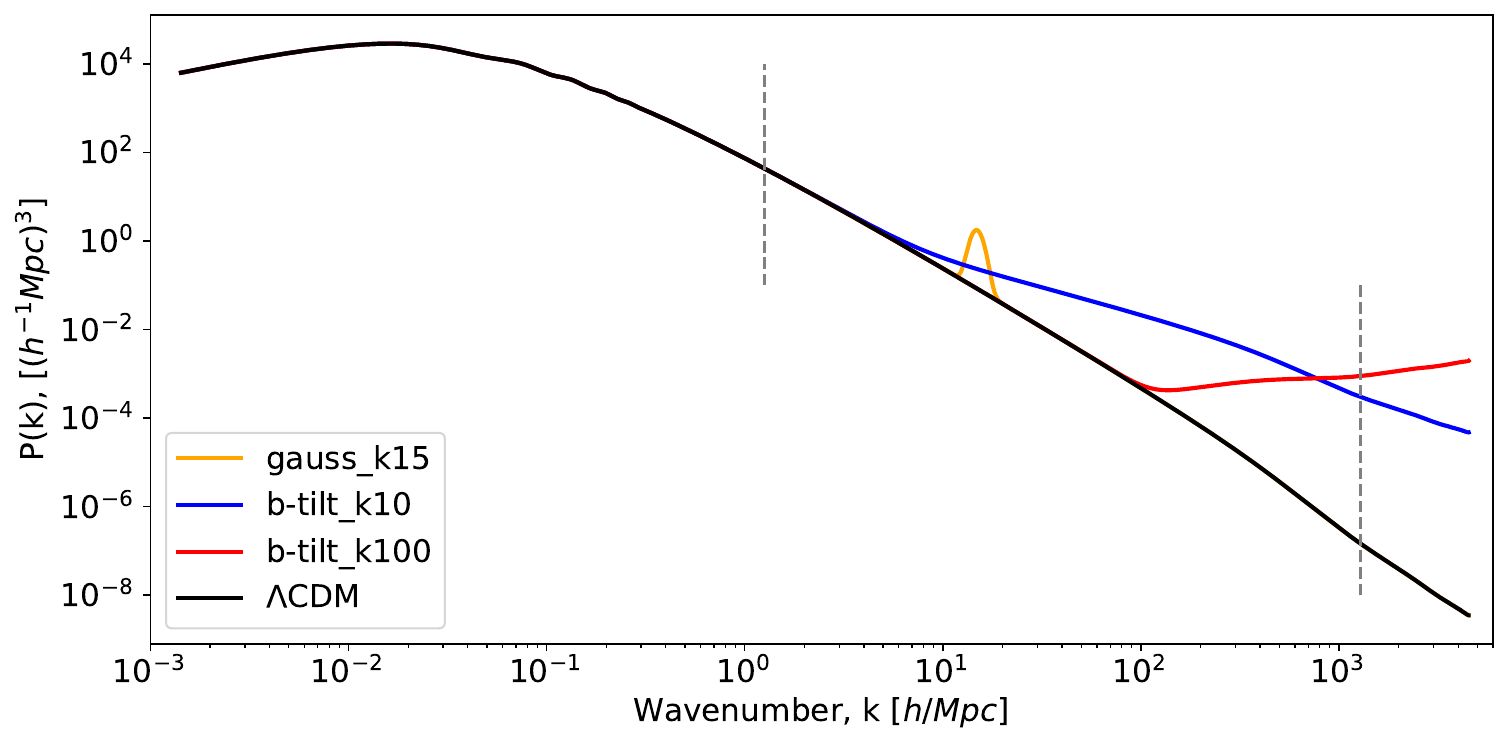}
    \caption{\textit{Top panel}: Transfer functions, applied to the 
    $\Lambda$CDM spectrum. \\
    \textit{Bottom panel:} resulting matter power spectra used in our simulation suite. \\
    On both panels grey dashed lines represent the Nyquist frequency for $(5\, $Mpc$/h)^3$ cube and ${2048}^3$ particles resolution. The parameters of each individual spectrum can be found in the Table~\ref{tab:sim}.
    }
    \label{fig:spectra}
\end{figure}

Additionally, the transfer function for tilted spectra is calculated as follows:
\begin{equation}
T(k) = \sqrt{1 + \frac{1}{p}\left(\frac{k}{k_0}\right)^{2p+2} },
    \label{tilt}
\end{equation}
where $p$ is constant. The transfer function essentially defines a smooth transition from the constant $T = 1$ to a power-law function $T(k) = k^{p+1}$, where the parameter $k_0$ defines the value of wave number where the transition happens (either 10 or 100, in our case).
In the Table~\ref{tab:sim} we provide the most relevant parameters of our simulations, including the values of constants from the eqs.~(\ref{bumps}) and (\ref{tilt}). Figure~\ref{fig:spectra} illustrates shapes of modifications.

\begin{table*}
\caption{Most relevant parameters of the simulation suite.} 
\centering
\begin{tabular}{p{0.30\textwidth}p{0.15\textwidth}p{0.15\textwidth}p{0.15\textwidth}p{0.15\textwidth}}
\hline
Main suite:                & \texttt{$\Lambda$CDM}  & \texttt{gauss\_k15}   & \texttt{b-tilt\_k10}   & \texttt{b-tilt\_k100}    \\ \hline
Box size $($Mpc$/h)$         & 5.0                    & 5.0                   & 5.0                  & 5.0                    \\
zoomed region resolution   & $2048^3$               & $2048^3$              & $2048^3$             & $2048^3$               \\
Initial redshift           & $300$                  & $1000$                & $1500$               & $1500$                 \\
Final redshift             & $8$                    & $8$                   & $8$                  & $8$                    \\
$k_0$                      & --                     & 15                    & 10                   & 100                    \\
$A$                        & --                     & 20                    & --                   & --                     \\
$p$                        & --                     & --                    & 0.5                  & 2.6                    \\
\hline
\end{tabular}
\label{tab:sim}
\end{table*}

We have run a series of four dark matter only simulations, employing the zoom technique to achieve high resolution in a specific region of interest. Each simulation utilizes a box size of $(5\,$Mpc$/h)^3$. Within the zoomed region, the resolution corresponds to $2048^3$ particles, while the intermediate levels are represented by $256^3$, $512^3$, and $1024^3$ particles, respectively. Three of the simulations utilize modified matter power spectra, while one employs the standard $\Lambda$CDM spectrum for comparison.

The simulations were run using the publicly available N-body code \texttt{GADGET-2} \citep{gadget}, which is widely used for cosmological simulations. This code utilizes a combined Tree + Particle Mesh (TreeMP) algorithm to calculate gravitational accelerations for each particle by decomposing the gravitational forces into a long-range term and short-range term interaction. Notably, \texttt{GADGET-2} is designed for MPI parallelization, which results in faster execution and scalability, allowing the code to handle a large number of particles with reasonable computational resources.

To account for the potential early formation of virialized structures, the simulations with tilted spectra start at $z = 1500$, simulation with the Gaussian bump spectrum starts at $z = 1000$, while the $\Lambda$CDM simulation starts at $z = 300$. The final redshift for all simulations is set to $z = 8$. This choice aims to minimize potential artifacts arising from the space periodicity of initial conditions within the relatively small simulation box.

Initial conditions for the simulations are generated using the publicly available code \texttt{ginnungagap} \footnote{https://github.com/ginnungagapgroup/ginnungagap}. The matter power spectrum for each simulation is defined individually by applying the appropriate transfer function. For the $\Lambda$CDM simulation, the power spectrum is generated using the publicly available code CLASS \citep{CLASS}. Importantly, the same initial random seed number is used for all simulations, ensuring that they differ solely in the amplitude of the power spectrum. Additionally, the amplitude of the longest wavelength mode in the generated initial conditions falls within 20\% of the theoretical value, mitigating the impact of cosmic variance on the high-mass end of the HMFs.

For each simulation, 100 snapshots are stored at redshift intervals equally spaced in logarithmic scale, spanning from $z = 25$ to $z = 8$. Halo analysis is subsequently performed using the publicly available code \texttt{AHF} \citep{AHF}. This analysis assumes that each halo comprises at least 5000 particles employs a virial overdensity criterion of 200 $\rho_{crit}$ and spatial resolution of the grid is limited to $5/2^{18}$Mpc$/h$. We also performed a similar analysis for the case where halo consisted of 50 particles (which is a default setting for \texttt{AHF}) and found no significant differences, therefore 5000 was taken as a more lightweight option. 

All simulations share the same cosmological parameters in agreement with  the values obtained by the \cite{planck}, i.e. $\Omega_m=0.31$, $\Omega_{\Lambda}=1-\Omega_m=0.69$, $\Omega_b=0.048$, $h=0.67$, $n_s=0.96$.

\section{Halo and subhalo mass functions\label{hsMF}}

\begin{figure*}
\includegraphics[width=0.9\textwidth]{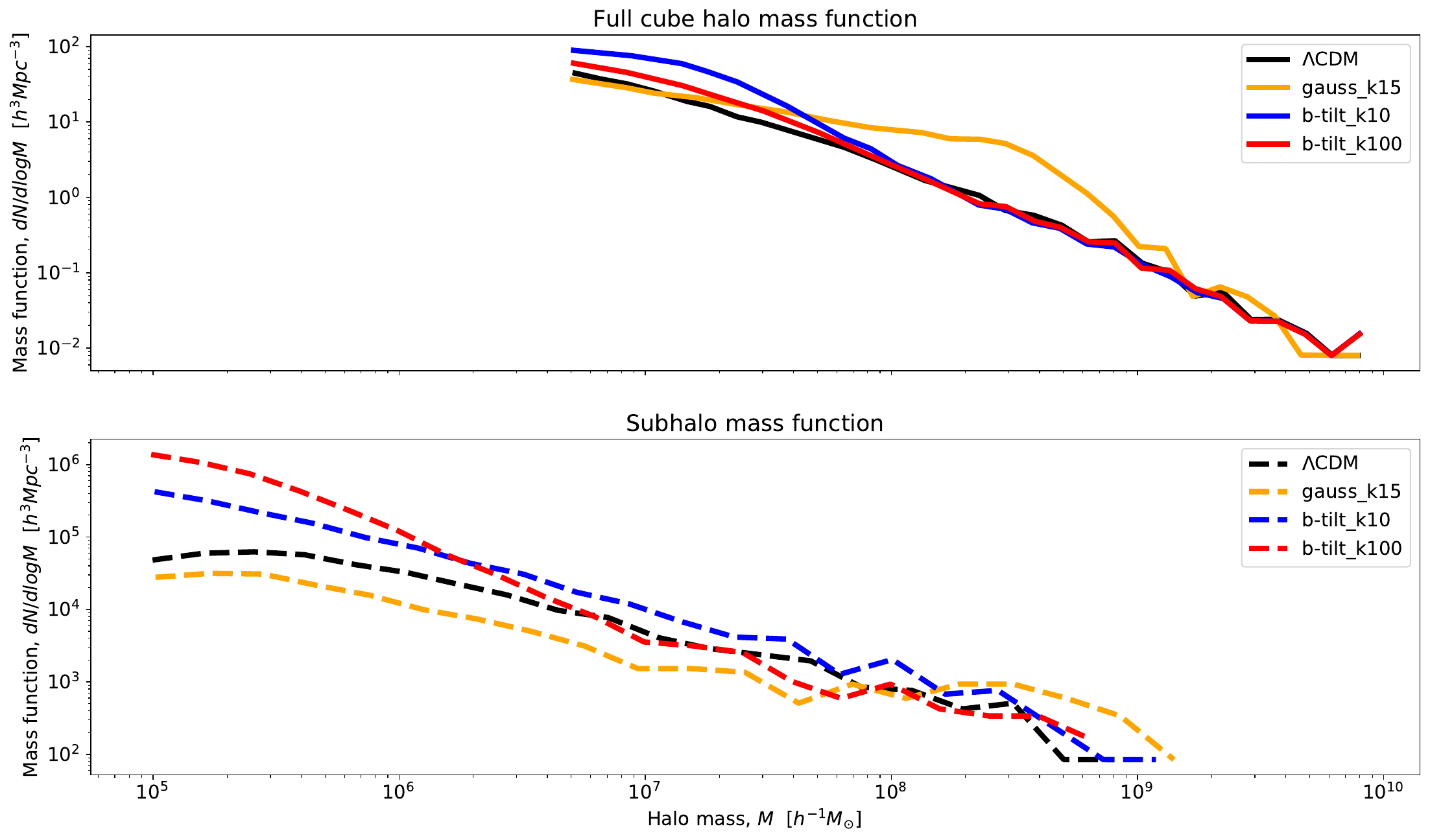}
\caption{\textit{Top panel}: Full-cube halo mass functions for the $\Lambda$CDM model and three models with modified spectra for the non-zoomed simulation with the resolution corresponding to $512^3$ particles.\\
\textit{Bottom panel}: Subhalo mass functions for the same models for zoomed simulation, described in the current paper. Each subhalo mass function is calculated for the 15 most massive halos in the zoomed region.}
    \label{fig:hmfs}
\end{figure*}

To investigate the impact of non-power-law primordial spectra on the halo abundance, we analyze the Halo Mass Functions (HMFs) obtained from our simulations.
Figure \ref{fig:hmfs} showcases the HMFs for the full simulation box (upper panel) and for subhalos of the 15 most massive halos in the zoomed region (\textit{bottom panel}). The upper panel displays the full-cube HMFs for the $\Lambda$CDM model and the three models with modified spectra, while the \textit{bottom panel} shows the subhalo mass functions for the same models.
Full-cube HMFs were obtained for the simulations with $512^3$ particles resolution (and other parameters identical to the zoom simulations presented in Section~\ref{description}). $\Lambda$CDM and  \textit{gauss\_k15} full box simulations were used in our previous work (see \cite{tkachev23}).

The HMFs for the $\Lambda$CDM model and the three models with modified spectra exhibit noticeable differences from each other. Particularly the \textit{b-tilt} models have significantly more amount of low mass halos in comparison with the $\Lambda$CDM. In the \textit{gauss\_k15} model there is an excess of halos in the range $3\times10^7<M<10^9$~M$_\odot/h$. This difference for the \textit{gauss\_k15} and $\Lambda$CDM models was discussed in detail in  \cite{tkachev23}.

For the subhalo mass function we used the zoomed simulations described in Section~\ref{description}. We selected the 15 most massive halos in the zoomed region and calculated the number of subhalos within those halos, then divided it by the total volume within the virial radii of the 15 most massive halos. All halos and subhalos were identified with \texttt{AHF}.
Naturally, the full-cube HMFs display significantly lower halo number densities, compared to subhalo mass functions, since the zoomed regions are placed around the most massive halos, leading to a higher matter density within those regions, compared to the full simulation box. 
Some effects of a local density increase (abundance of subhalos in and around massive halo) on the halo mass function in cosmological models with a bump were recently considered in \cite{eroshenko}. Earlier, such an influence was studied for models with power law spectra in \cite{arkhipova}.

Interestingly, the subhalo mass function for the \textit{gauss\_k15} model demonstrates a less prominent excess above $\Lambda$CDM (around $M \simeq 3\times10^7-10^9 M_{\odot}/h$), compared to the full-cube halo mass function for the same model. On the other hand, the \textit{b-tilt\_k10} model demonstrates an overall increase in the abundance of subhalos, while \textit{b-tilt\_k100} model shows a more pronounced enhancement of high-mass halos, notably for subhalos. 
Furthermore, the trend found in the subahlo abundance (particularly for \textit{b-tilt\_k10} model) is largely consistent with the results from Esteban et al. \citep{Esteban}, who studied a modified DM power spectra in the context of local Galaxy Group structure.

These findings suggest that modifications to the primordial power spectrum can have a significant impact on the abundance of halos and subhalos.
We need to mention that the accurate treatment of subhalo survival to $z=0$ requires special simulations with properly chosen softening length, otherwise they may be subject to artificial disruption (see, e.g. \cite{subhalo1,subhalo2}). Our simulations were made with the focus on the field halos at large redshifts, so our softening length choice is not optimal for subhalos. Thus, our results on them could suffer from numerical effects, especially for the smallest subhalos produced in the models with the blue tilted power spectrum.
Further analysis of these differences could shed more light on the role of small-scale power in the formation and evolution of halos.

\section{\label{denprofiles}Density profiles}

The universality of the NFW profile $\rho(r) = \rho_0(r_s/r)(1 + r/r_s)^{-2}$ has been discussed for many years \cite{uni_nfw1, uni_nfw2, uni_nfw3, 1997ApJ...490..493N, 2001MNRAS.321..559B, 2003MNRAS.341.1311S, 2010MNRAS.402...21N}.  
Numerical simulations provided under different assumptions result in the same shape independent on them. However, the class of non-power-law spectra of density perturbations has not yet been studied. 

\begin{figure}
\hbox{\hspace{-0.9em} \includegraphics[width=1.0\linewidth]{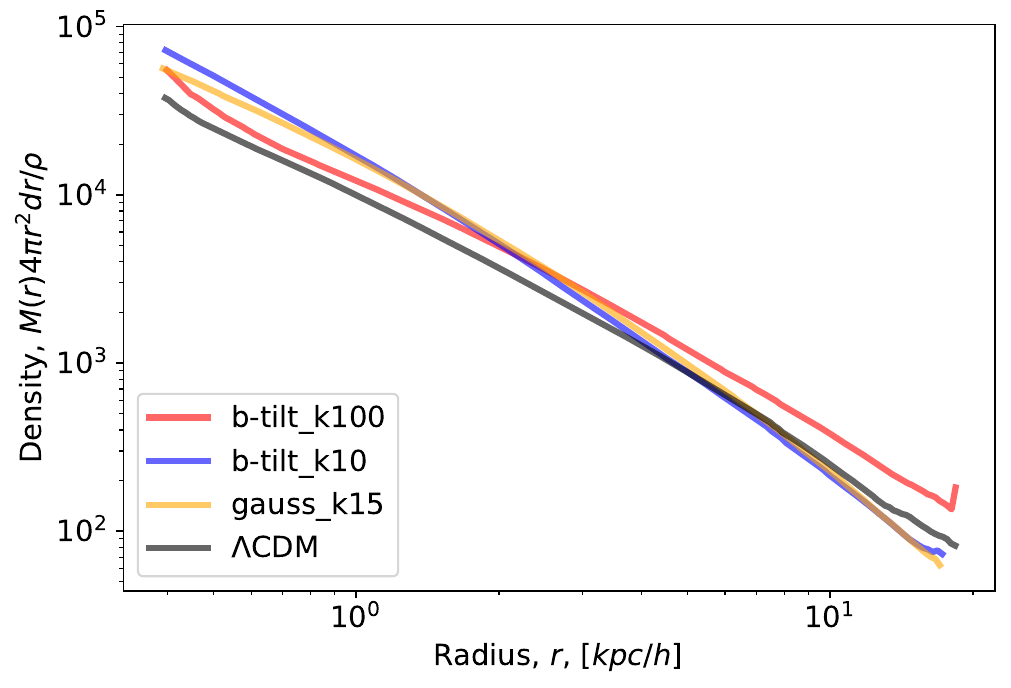}}
\hbox{\hspace{1.0em}\includegraphics[width=0.94\linewidth]{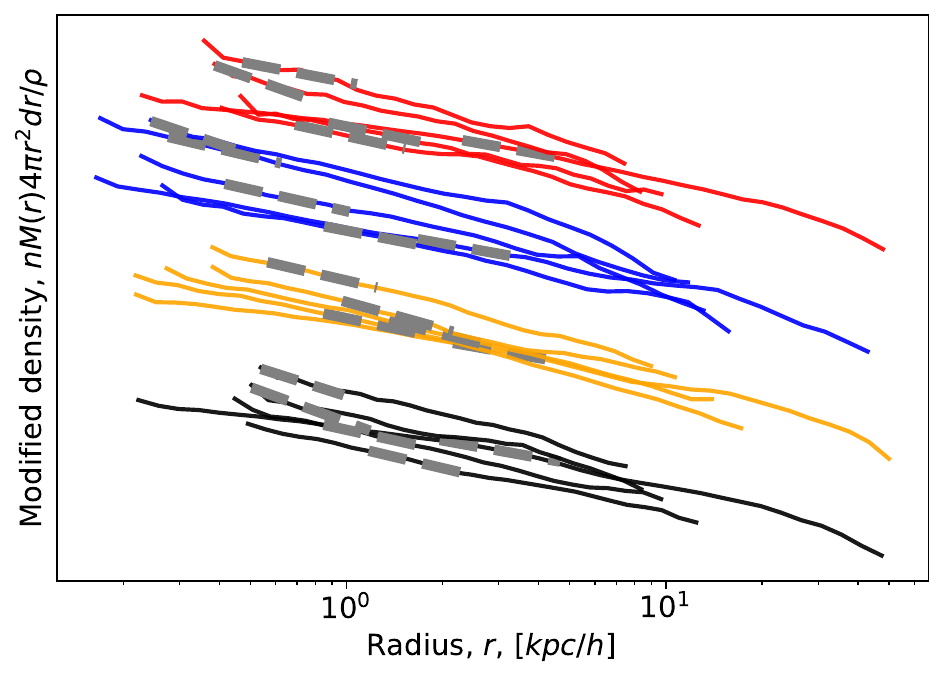}}
\caption{\textit{Top panel:} An example mean density power profiles, averaged for approximately 100 halos at mass range $M \simeq 10^8-10^9 M_{\odot}$ at redshifts between $z \simeq 9-10$.\\
\textit{Bottom panel:} An example power-law approximation for density profiles at $z=8$ for the part of the halo where $\frac{1}{5}r_s < r < \frac{1}{2}r_s$. The profiles are artificially shifted vertically in order to increase visibility. Each model is represented by the different color (as shown in legend), and the power-law fit is depicted as a \textit{dashed gray line}.
}
\label{fig:profsapprox}

\end{figure}

To investigate the impact of non-power-law primordial spectra on the internal structure of dark matter halos, we analyze the density profiles obtained from our simulations. Figure~\ref{fig:profsapprox} showcases the mean density profiles for halos with a mass $M \simeq 10^8-10^9 M_{\odot}$, which roughly corresponds to the mass range where the difference between halos from $\Lambda$CDM model and modified spectra models should be the most significant due to extra power at redshifts between $z=9$ and $z=10$. While the profiles exhibit similarities, subtle differences emerge between the various models, particularly in the inner regions. This suggests that modifications of the primordial power spectrum can indeed influence the central density distribution within dark matter halos.

To quantify these differences, we employ power-law approximations for the density profiles, as exemplified in the \textit{bottom panel} of Figure~\ref{fig:profsapprox}  for the several most massive halos at redshift 8. However, fitting the central part of the halo profile (as was done in e.g. \cite{Ricotti_2003}) might not be sufficient, as was shown by \cite{Knollmann_2008}, since the slope of the profile depends on the halo concentration, and is shallower for less concentrated halos. Therefore, we attempt to eliminate this bias by fitting the power-law only to a specific radius range, where the concentration of different models is supposed to be similar, i.e., we fit a power-law function only to the part of the halo where $\frac{1}{5}r_s < r < \frac{1}{2}r_s$. Here $r_s$ is the scale radius of the halo and can be calculated as $r_s = R_{vir}/c$, where $R_{vir}$ is the halo radius and $c$ is a halo concentrations (as defined in \cite{Prada_2012}). For NFW halo $r_s$ indicates the transition between the central and the peripheral regions of halo, therefore, the choice of such radius range suggests that we focus on the central regions. Unfortunately, we can not set the range significantly lower than $\frac{1}{5}r_s$, as our resolution does not allow that. Although, as can be seen from the \textit{bottom panel} of the Figure~\ref{fig:profsapprox}, for each halo the range of approximation is slightly different, and for heavier halos it is shifted towards the periphery.

Once we have established the methodology for fitting the power-law to the halo profiles from our simulations, we also compare each profile with the corresponding NFW profile, using the calculated $r_s$ values for each given halo, while $\rho_0$ is calculated from the integrated virial mass $M_{vir}$ (which we take as the mass of the halo) within the virial radius of the halo $r_{vir}$:
\begin{equation}
M_{vir} =  4 \pi \rho_0 r_s^3 \left[ \ln \left( \frac{r_s + r_{vir}}{r_s} \right) - \frac{r_{vir}}{r_s + r_{vir}} \right].
\end{equation}
Further we calculate the slope of the resulting NFW profile at the same radius range of $\frac{1}{5}r_s < r < \frac{1}{2}r_s$, which for NFW profile should not vary for the same fractions of $r_s$ (e.g. the slope of the NFW profile at $r = r_s$ should be equal to -2 by definition).

\begin{figure*}
\includegraphics[width=0.9\textwidth]{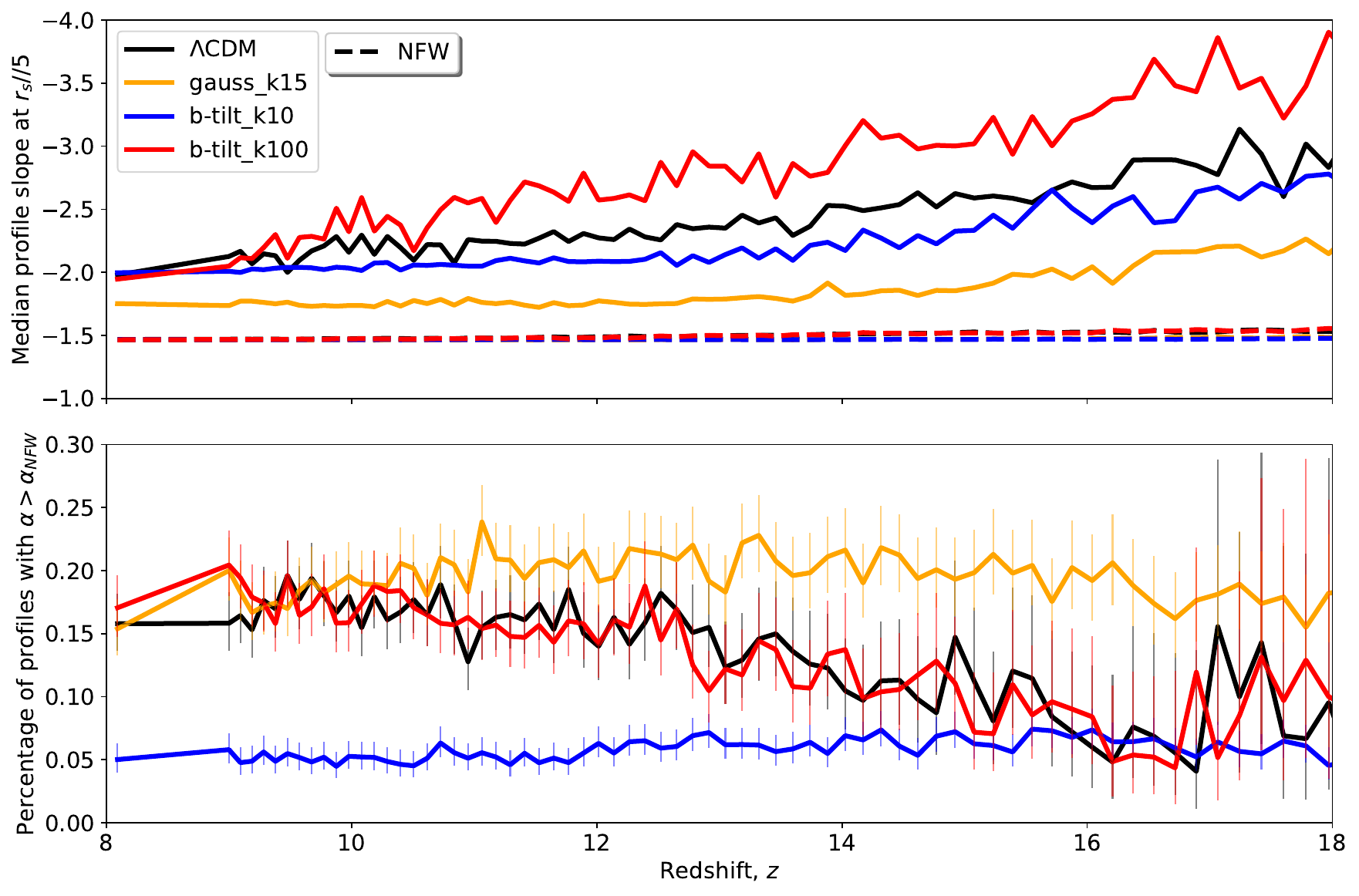}
\caption{\textit{Top panel:} Evolution of the median slope $\alpha$ of the halo profiles (\textit{solid lines}) at range $\frac{1}{5}r_s < r < \frac{1}{2}r_s$ as a function of redshift $z$. The \textit{dashed lines} show the median slope for the NFW profiles as a function of redshift $z$.\\
\textit{Bottom panel:} Percentage of halo profiles from our simulations that have a slope $\alpha>\alpha_{NFW}$ less steep than the NFW profile in the same radius range $\frac{1}{5}r_s < r < \frac{1}{2}r_s$.}
    \label{fig:inc_hist}
\end{figure*}

The \textit{top panel} of the Figure~\ref{fig:inc_hist} shows the evolution of the median slope $\alpha$ of the halo profiles (\textit{solid lines}) at range $\frac{1}{5}r_s < r < \frac{1}{2}r_s$ as a function of redshift between $z \simeq 8$ and $z \simeq 18$. The \textit{dashed lines} show the median slope for the NFW profiles at the same radius range -- which, as expected, remains constant $\alpha_{NFW} \simeq -1.5$.
Additionally, the \textit{bottom panel} of the Figure~\ref{fig:inc_hist} shows the percentage of halo profiles from our simulations that have a slope $\alpha>\alpha_{NFW}$ less steep than the according NFW profile in the same radius range $\frac{1}{5}r_s < r < \frac{1}{2}r_s$.

The panels indicate that at higher redshifts the median slope of the profiles for all models (including $\Lambda$CDM) is significantly steeper than for the according NFW profile. On the other hand, compared to $\Lambda$CDM model, the median profile slopes behave differently for different modified spectra models, such as the Gaussian bump model $gauss\_k15$ has profiles with significantly smaller slopes, while the model \textit{b-tilt\_k100} demonstrates significantly higher halo slopes.
The percentage of halos exhibiting shallower profiles than the NFW profile also differs for most models, such as for \textit{b-tilt\_k10} and $gauss\_k15$ models it remains constant at approximately $5\%$ and $20\%$ respectively, while for $\Lambda$CDM and \textit{b-tilt\_k100} models it increases with time from $5\%$ and $20\%$.
At smaller redshifts these differences between models become smaller and almost disappear at $z \simeq 8$, but \textit{b-tilt\_k10} stays alone.

\begin{figure*}
\includegraphics[width=0.98\textwidth]{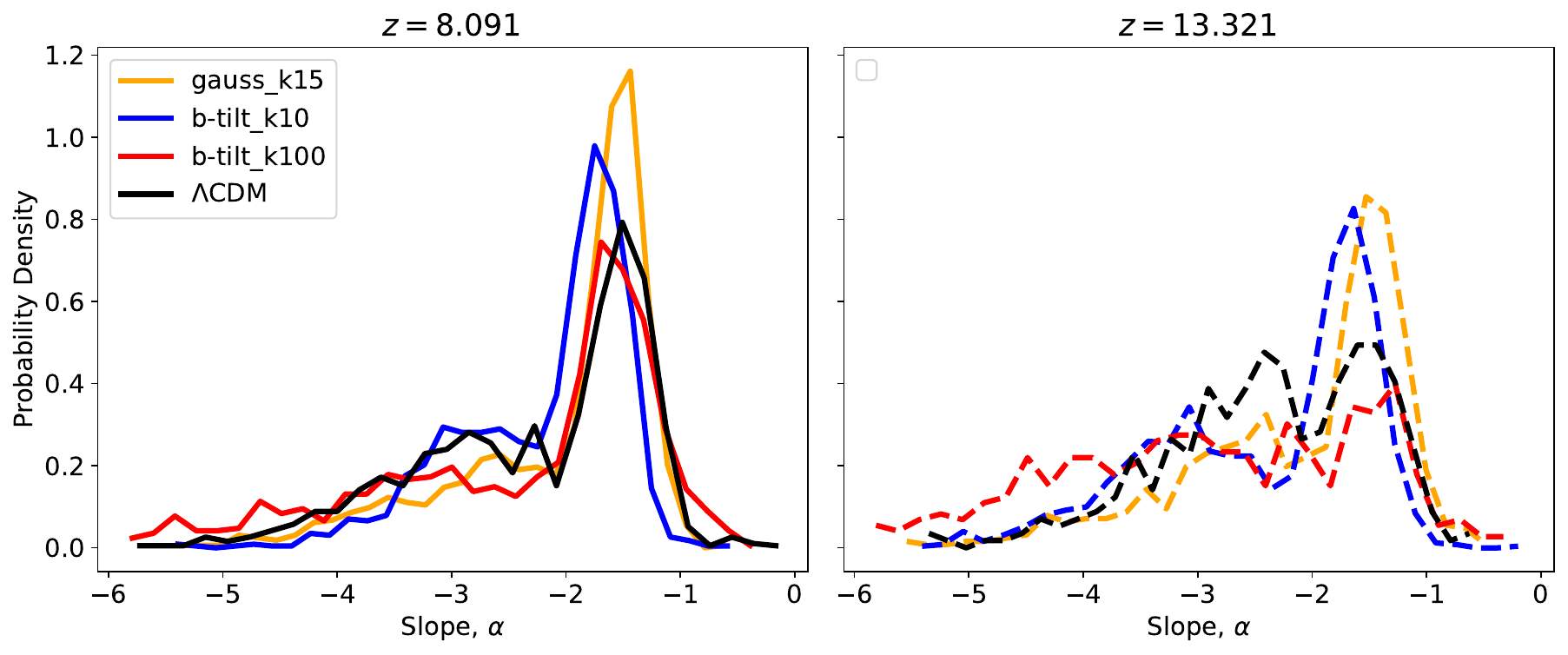}
\includegraphics[width=0.98\textwidth]{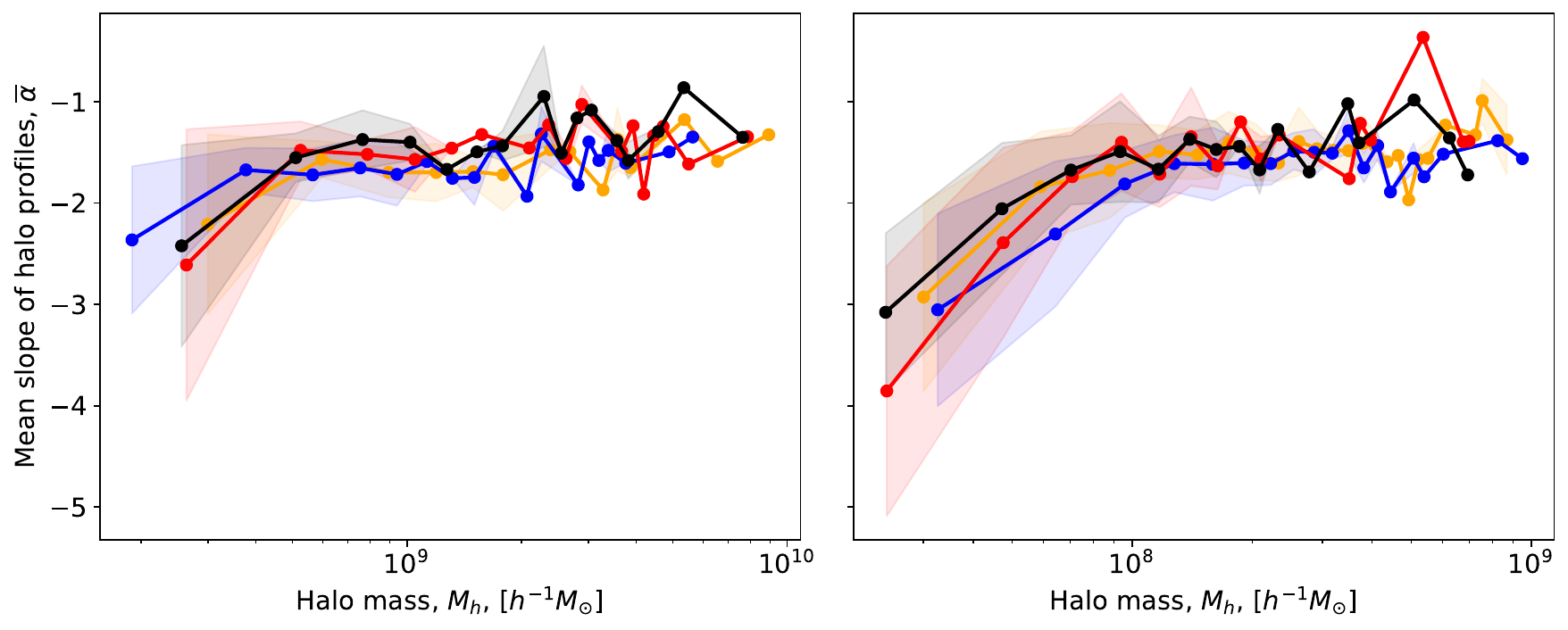}
\caption{\textit{Top panel:} Density distribution of the slope of halo profiles for different models at redshifts $z = 8.091$ and $z = 13.321$, consequently.\\
\textit{Bottom panel:} Dependence of the halo profile slope from mass of the halo at the same redshifts. Colored areas represent the $1\sigma$ value level of dispersion.}
    \label{fig:inc_mass}
\end{figure*}

This behavior becomes more apparent if we look at the according distributions of slopes for different models at redshifts $z = 8.091$ and $z = 13.321$, as displayed by \textit{top panel} of Figure~\ref{fig:inc_mass}. The distributions exhibit quite significant (and varying) left-side tails, while at smaller redshifts the tails decrease and all 4 models start demonstrating almost identical distributions.
Additionally, the \textit{bottom panel} displays the dependence of halo profile slope from mass of the halo. As can be seen, all the simulations on both panels exhibit approximately the same trend, where smaller mass halos generally have steeper profile slopes, but at higher redshifts the dispersion for smaller mass halos is significantly higher.

These findings challenge the universality of the NFW profile and highlight the sensitivity of halo structure to the shape of the primordial power spectrum. Modifications to the power spectrum, such as the Gaussian bump and tilted spectra explored in this work, can lead to significant deviations in the inner density profiles of dark matter halos, particularly for less massive halos and at earlier epochs of the Universe.

\section{\label{compensatio}Compensation of the small scale power cut}
In this Section we consider the impact of the small scale power which was cut from the simulation due to the limited resolution of the initial conditions. The dimensionless power spectrum of density perturbations, $\Delta(k) \equiv k^3 P(k)$ behaves as $\sim \log(k)$ at large $k$ until the free streaming scale. This scale is usually not resolved in cosmological simulations, so the density perturbations below the Nyquist scale $k_{Ny} = \pi/(L_{box} N_{1D})$ are missing. Since $\Delta(k)$ grows towards large $k$, these missing perturbations should have became nonlinear earlier than the perturbations resolved by the simulation. As has been proposed in \cite{DLM2007, DLM2008, LM2010} the missing perturbations could generate additional entropy which could destroy cusps.

We try to emulate the effect of missing perturbations on density profiles in several ways. First of all, one can add power at small scales resolved by the simulation. This actually was done in our simulations with various non-standard power spectra described in Section~\ref{description}. However, as was shown in Section~\ref{denprofiles}, this has not lead to a significant flattening of the cusps. However considering the proposal of \cite{DLM2007, DLM2008, LM2010}, one also could directly add the entropy produced by missing perturbations in a form of random velocities. 

Let us briefly remind the proposition of \cite{DLM2007, DLM2008, LM2010}. There an entropy function was introduced:
\begin{equation}
    E = \sigma^2 n^{-2/3},
\end{equation}
where $\sigma$ is the velocity dispersion in some volume and $n$ is the particle number density in that volume. From hydrostatic equilibrium equation one can easily find that for the cusp $E \rightarrow 0$ as $r \rightarrow 0$ while for the constant density core $E \rightarrow const$. Since $E$ is a function of the coarse-grained entropy, we expect $E$ can only grow when a halo forms. In CDM model initial flows of matter are cold on the linear stage, $E \rightarrow 0$ as the considered volume shrinks to zero size. However, nonlinear evolution can ''thermalize'' linear velocity perturbations and produce nonzero $E$. We make simulations of this effect by adding these thermal velocities by hand.

\begin{figure*}
\includegraphics[width=0.49\textwidth]{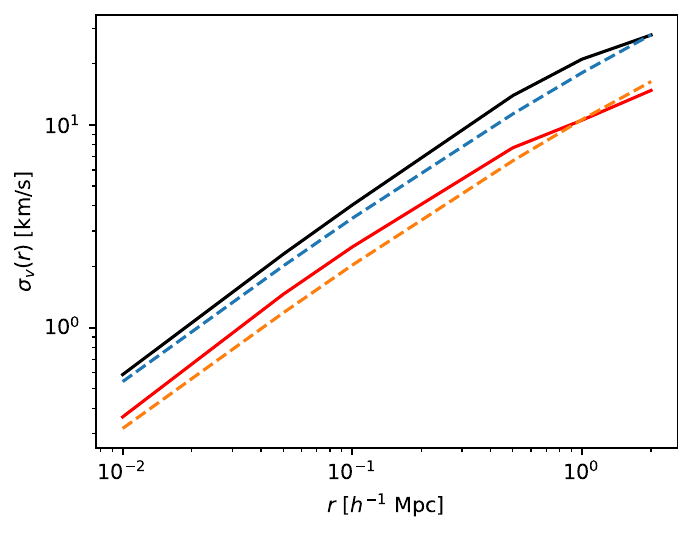}
\includegraphics[width=0.49\textwidth]{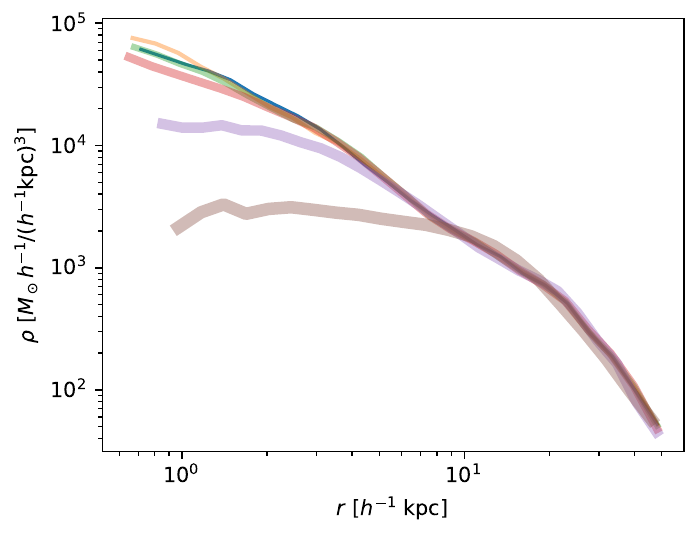}
\caption{\textit{Left panel:} the velocity dispersion in a sphere of radius $r$ at $z=25$ (two bottom curves) and $z=8$ (two upper curves). Solid lines represent measurements from the $\Lambda$CDM simulation while dashed lines show linear theory predictions.\\ 
\textit{Right panel:} Density profiles at $z=8$ for the most massive halo in simulations with artificially added velocities of $\sigma_v=0.5$, 1.0, 2.0, 4.0, 8.0 and 16.0~km/s at $z=25$. The thicker the line, the higher the velocity dispersion.}
    \label{fig:sigma_v}
\end{figure*}

The amount of velocities can be estimated using the linear theory. We assume that when a particular scale becomes nonlinear, the particle velocities are randomized but have amplitudes in accordance with the linear theory. This is supported by measurements of the velocity dispersion in the $\Lambda$CDM simulation shown in Fig.~\ref{fig:sigma_v}, left panel. In that Figure, velocity dispersions are shown at two different redshifts. At $z=25$ all the shown scales are in linear regime and the simulation data almost coincides with the linear theory calculation. At $z=8$ all the plotted scales are in the non-linear regime, but the velocity dispersion is still well described by the linear theory. According to this argument, the Nyquist scale in our simulation goes nonlinear at $z=25$ and linear theory velocities at this scale and time have amplitude of $\sigma_v = 0.5$~km/s.

To test the impact of random velocities on density profiles we run a set of simulations with random velocities added at the simulation snapshot at $z=25$. One should note that these random velocities evolve like a decaying mode of perturbations, so they decrease with time as $(1+z)$. Fig.~\ref{fig:sigma_v}, left panel, shows that velocities do not decline after perturbations getting nonlinear. If we consider the last snapshot of our simulation, at $z=8$, the amplitude of linear theory velocities at Nyquist scale is 1.5~km/s, and if we take into account the decay of random velocities, we should set $\sigma_v=4.8$~km/s at $z=25$ to get 1.5~km/s at $z=8$. This is the maximal velocity dispersion which can be achieved in the frame of this ''lost entropy'' proposal of \cite{DLM2007, DLM2008, LM2010}. Since this behavior of decaying mode results in some uncertainty in the initial velocity dispersion which is needed to compensate the lost small-scale perturbations, we set $\sigma_v=0.5$, 1.0, 2.0, 4.0, 8.0 and 16.0~km/s at $z=25$ and check how it affects the density profiles of halos.

The density profiles obtained in these random velocity simulations are shown in Fig.~\ref{fig:sigma_v}, right panel. One can see that adding random noise with $\sigma_v = 8$~km/s indeed results in the flattening of the cusp, however this velocity is higher than the estimate obtained in theory. These high velocities also produce another visible effect: damping of small scale density perturbations, and, as a result, significant decrease of the number of low massive halos. This is a side effect which should not be present in an ideal simulation with ''infinite'' resolution.

We conclude that the addition of random velocities to particles allow to destroy cusps in density profiles, however the amplitude of the velocities needed for this is higher than expected to compensate the missing small scale power, and also such ''compensation'' results in an artificial destruction of small halos.

These additional velocities should not be mixed up with the warm DM models. In those models DM particles also have random thermal velocities in the early universe, but these velocities decay well before the halo formation starts, leaving a cold flow of particles with suppressed small scale perturbations. As we have pointed in the Introduction, simulations with suppressed small scale power show that cusps are retained in halos in these models. Additional velocities at the right time can arise in exotic DM models whith decaying DM particles if the decay produces a kick on the particle remnant with the right amplitude and at the right time, which requires fine tuning of such a model, see \cite{Pilipenko09}.

\section{\label{conclusions}Conclusions and Discussions}
We studied the impact of the primordial spectrum on the density profile of gravitationally bound DM halos. All considered models have some enhancement on a scale less than $1\,h^{-1}$~Mpc, but it was realized in different ways. One spectrum had a bump at $k_0=15\,h$~Mpc${}^{-1}$, and two spectra are blue-tilted with characteristic scales $k_0=10\,h$ and $100\,h$~Mpc with small-scale additional (to standard $\Lambda$CDM) slopes 1.5 and 3.6, consequently. More detailed description of considered models can be found in eqs.~(\ref{bumps})-(\ref{tilt}), the Table~\ref{tab:sim}, and the Figure~\ref{fig:spectra}.  

We also investigated how modified primordial spectra affects the abundance of halos and subhalos. Our analysis revealed significant differences in the halo mass functions (HMFs) and subhalo mass functions for various models, particularly for lower halo masses in the case of blue-tilted spectra and for higher halo masses in the case of the \textit{gauss\_k15} model. Additionally, compared to the standard HMFs, the subhalo mass function for the \textit{gauss\_k15} model demonstrates a less prominent bump, while the \textit{b-tilt\_k10} model shows an overall increase in subhalo abundance.

We analyzed the evolution of individual and averaged density profiles for the redshift interval from $z=18$ till $z=8$ (the latter was the final value in our N-body simulation) and found out that the median profile slopes in all models (including standard $\Lambda$CDM one) are steeper in comparison to the NFW profile. The Figure~\ref{fig:inc_hist} demonstrates that density profiles of all cosmological models vary with redshift $z$ from cuspy values in the interval from $\alpha\in(-3;-2)$ to $-1.5$ which is just the value for the slope of NFW profile at $r_s/5$ (dashed lines). Despite these slopes corresponding to a cusp in the inner part of a halo, the percentage of density profiles with a shallower slope $\alpha>\alpha_{NFW}$ is close to $5\%$ for \textit{b-tilt\_k10} model and to $15 - 20\%$ for the rest of models, e.g. we detect in simulations some number of cored halos. As to evolution of the percentage with redshift, it can be considered as negligible due big uncertainties at high $z$.

Looking for the $z$-evolution of probability density of finding a halo with some slope value, we find that it becomes sharper at smaller redshift showing a tendency to unify the density profiles of halos.

We also study how the variety of profile slopes appears for different halo masses and despite the significant difference in initial conditions at different redshifts.  The figure~\ref{fig:inc_mass} demonstrates that the low-massive tail of halos ($M>10^8M_\odot$) at $z\simeq 13$ has a more cuspy slope then the more massive one and for smaller redshift.

To clarify a possible way to solve the cusp problem by enhancing a matter spectrum, we also considered a toy model with boosted small scale random velocities. We find that it results in the abundant generation of cored halos accompanied by suppression of sub-halos. It could be considered as an elegant solution of the \textit{too-big-to-fail} problem too, but this solution has a high price, which is a rather high value of the velocity. 


\begin{acknowledgments}
The work was supported by the Russian Science Foundation (grant number 23-22-00259). We are grateful to the anonymous referee for their feedback, which has helped us to clarify the text of our paper. 
\end{acknowledgments}


\newcommand{\mnras}{MNRAS}
\newcommand{\jcap}{J. Cosmology Astropart. Phys.}
\newcommand{\apjs}{ApJS} 
\newcommand{\aap}{A\&A} 
\newcommand{\apjl}{ApJ Letters}

\bibliography{refs}

\end{document}